\documentclass[%
 reprint,
superscriptaddress,
amsmath,amssymb,
aps,
]{revtex4-2}

\usepackage{graphicx}
\usepackage{dcolumn}
\usepackage{bm}
\usepackage{hyperref}
\usepackage{verbatim} 
\hypersetup{
colorlinks,
citecolor=blue,
filecolor=blue, 
linkcolor=blue, 
urlcolor=blue
}

\usepackage{color}
\usepackage{xcolor}
\usepackage{ulem}
\usepackage{physics}
\usepackage{multirow}

\newcommand{\q}{{\mathbf q}}

\DeclareMathAlphabet{\mathgothic}{U}{euf}{m}{n}
\renewcommand{\Re}{\mathgothic{Re}}
\renewcommand{\Im}{\mathgothic{Im}}

\newcommand{\editor}[2]{%
  \expandafter\newcommand\csname #1note\endcsname[1]{%
    \textcolor{#2}{(\textbf{#1:} \textit{##1})}}%
  \expandafter\newcommand\csname #1\endcsname[1]{%
    \textcolor{#2}{##1}}%
  \expandafter\newcommand\csname #1cancel\endcsname[1]{%
    \textcolor{#2}{\sout{##1}}}%
\expandafter\newcommand\csname #1can\endcsname[1]{%
    \textcolor{#2}{\sout{##1}}}%
  \expandafter\newcommand\csname #1change\endcsname[2]{%
    \textcolor{#2}{\sout{##1} ##2}}%
      \expandafter\newcommand\csname #1ch\endcsname[2]{%
    \textcolor{#2}{\sout{##1} ##2}}%
  \newenvironment{#1text}{\color{#2}}{\color{black}}
}
\definecolor{Blu}{rgb}{0.00,0.00,1.00}
\definecolor{Red}{rgb}{1.00,0.00,0.00}
\definecolor{Cyan}{rgb}{0.00,0.50,0.50}
\definecolor{Green}{rgb}{0.00,0.70,0.00}

\editor{DAL}{Green}
\editor{R}{Red}

\renewcommand{\emph}{\textit}

\begin{document}

\title{Predicting plasmonic pole representations of complex dielectric spectra using convolutional neural networks}

\author{Dario A. Leon}
\email{d.a.l.valido@smn.uio.no}
\affiliation{Centre for Materials Science and Nanotechnology, University of Oslo, 0318 Norway}%

\date{\today}

\begin{abstract}
Extracting spectral properties such as energy position, broadening, and spectral weight from dielectric spectra is critical for interpreting collective electronic excitations in many-body physics. Conventional approaches typically rely on nonlinear fitting procedures that can be computationally demanding and highly sensitive to initialization and fitting choices. In this work, we develop a convolutional neural-network framework for the direct inversion of dielectric spectra into their underlying plasmonic pole structures using a multipole-Pad\'e representation. Rather than training on a constrained database of spectra associated with a specific set of materials, the network is trained entirely on synthetic spectra generated from the analytical multipole-Pad\'e expression with randomized parameters. 
This enables the network to learn the general mapping between the spectra and their features in an unbiased way, without requiring large, material-specific datasets derived from real materials. We demonstrate that the synthetically trained network generalizes to complex first-principles and experimental dielectric spectra, extracting the underlying pole parameters with high accuracy in a single forward pass. This approach provides an efficient and robust alternative to conventional nonlinear fitting, enabling high-throughput, automated analysis of dielectric spectra across diverse materials and applications.
\end{abstract}

\maketitle

\section{Introduction}
Machine-learning (ML) methods have increasingly been used to accelerate the calculation and interpretation of spectroscopic and dielectric properties of materials. Neural-network and related approaches have been developed to predict dielectric response functions and optical spectra from material-specific information~\cite{Ward2016npjCompMat,Chen2020npjCompMat,Zhang2020PRB,Dong2021CS,Shimano2023SciRep,Grumet2024JPCC,Grunert2024PRM,Ibrahim2024AMI,Borlido2026PRM}, as well as to analyze and interpret experimental spectroscopic data~\cite{Stein2019CS,Anker2023CS,del-Pozo-Bueno2023Um,Gleason2024npjCompMat,Penfold2024MLsciandtech,Mizoguchi2026Micro}. 
A distinct class of inverse problems arises when the response function is itself available, but the physical parameters underlying its spectral structure must be inferred. Rather than predicting a response function from material information or experimental data, the task is to extract a physically meaningful representation of the underlying excitations.

The frequency-dependent inverse dielectric function, $\varepsilon^{-1}(\omega)$, provides a direct description of the dynamical screening response of an interacting electron system and plays a central role in the study of collective electronic excitations~\cite{Leon2026JPCM}. It enters, for example, in many-body approaches such as the $GW$ approximation and the Bethe-Salpeter equation (BSE)~\cite{martin2016book}, as well as in the interpretation of experimental probes including optical and absorption spectroscopy, inelastic x-ray scattering~\cite{Schulke2007book}, and electron-energy-loss spectroscopy (EELS)~\cite{Egerton2011book}. The analytical structure of $\varepsilon^{-1}(\omega)$ in the complex-frequency plane provides a compact representation of these excitations: collective modes appear as poles at complex frequencies, with the real part of a pole specifying its characteristic excitation energy, its imaginary part describing the associated broadening, and its residue determining the corresponding spectral weight. Constructing this pole representation from a numerical spectrum therefore provides a physically interpretable parametrization of the underlying dynamical response.

Approaches based on multipole-Pad\'e approximants (MPA) have been developed and applied in several contexts, including efficient $GW$ calculations and the analysis of optical, absorption, and EELS~\cite{Golze2019FC,Lee1996PRB,Jin1999PRB,Soininen2003JPCM,Leon2021PRB,Leon2023PRB,Guandalini2024PRB,Leon2025PRB,Leon2026PRB,Leon2026Arxiv,Bonacci2026JCTC}. In first-principles applications, the MPA representation can be constructed efficiently from a limited number of frequency samples~\cite{Leon2021PRB,Leon2023PRB,Leon2025PRB}, while fitting approaches can also be used when the full-frequency response is available~\cite{Leon2026PRB,Leon2026Arxiv}. Although these methods provide powerful tools for obtaining compact pole representations, constructing them across large numbers of spectra and materials can become computationally challenging, particularly when automation and high-throughput analysis are required. This creates an opportunity for ML methods to learn the inverse mapping from the frequency-dependent response to its underlying multipole structure.

In this work, we develop a convolutional-neural network (CNN) model for directly inferring the MPA representation of a frequency-dependent response. Rather than training on spectra associated with a particular set of materials, we construct a synthetic ensemble directly from the analytical MPA expression, with pole positions, broadening, and residues sampled over broad parameter ranges. This allows the CNN model to learn the general mapping between the spectral response and its underlying multipole structure, while avoiding a dependence on material-specific correlations in the training data. Once trained, the CNN model extracts the MPA parameters in a single forward pass, providing a rapid and automated alternative to traditional numerical methods. We demonstrate the approach first on synthetic spectra and subsequently on complex first-principles and experimental dielectric responses, including spectra containing multiple overlapping collective excitations.

The paper is organized as follows. Section~\ref{sec:theory} presents the theoretical framework, including the first-principles calculation of the dynamical dielectric response and its MPA representation. Section~\ref{sec:ML_framework} describes the ML framework, 
including the formulation of the inverse problem, construction of the synthetic ensemble, CNN architecture, and training objective. 
Section~\ref{sec:results} presents the results, including validation on synthetic spectra, the influence of spectral superposition and weak perturbations, and the application of the trained ML model to first-principles and experimental dielectric spectra. Finally, Sec.~\ref{sec:conclusions} summarizes the main conclusions.

\section{Theoretical Framework}
\label{sec:theory}

\subsection{Dynamical dielectric response}
\label{sec:dielectric_response}
The dielectric response describes the screening of the Coulomb interaction in an interacting electron system and can be formulated within many-body perturbation theory (MBPT) in terms of the density-response function. In a compact notation, the inverse dielectric function is written as
\begin{equation}
\begin{aligned}
\varepsilon^{-1}(\omega) &=1+Y(\omega), \\
Y(\omega) &\equiv v\chi(\omega),
\label{eq:epsilon}
\end{aligned}
\end{equation}
where $v$ denotes the Coulomb interaction and $\chi(\omega)$ is the interacting density-response function. The quantity $Y(\omega)$ therefore contains the dynamical contribution to the inverse dielectric response and is the quantity modeled throughout this work.

Within MBPT, the interacting response can be obtained from a Dyson equation that relates the full response to the irreducible polarizability. At the random-phase approximation (RPA) level, the irreducible polarizability is approximated by the independent-particle polarizability, and the resulting Dyson equation accounts for Coulomb screening processes. In this approximation, the density-response function satisfies
\begin{equation}
\chi(\omega)=\chi_0(\omega)
+\chi_0(\omega)v\chi(\omega),
\label{eq:dyson}
\end{equation}
where $\chi_0(\omega)$ is the independent-particle polarizability. 
The RPA approximation provides a physically transparent description of dynamical screening and collective electronic excitations and has been widely used to describe plasmonic excitations and energy-loss spectra (see e.g. Refs.~\cite{Leon2024PRB,Leon2026JPCM,Leon2026PRB}). 

In the present work, the first-principles spectra used as reference for a selected set of materials were calculated at the RPA level. This provides a well-defined and physically motivated benchmark for assessing the ability of the CNN model to infer the underlying multipole structure. Importantly, the synthetic training data are generated directly from the analytical MPA representation and are therefore independent of any particular level of many-body theory. The use of RPA spectra in the first-principles benchmark thus defines the reference data used to test the model rather than a theoretical constraint imposed on the learned inverse mapping. The inversion procedure can in principle be applied to responses obtained from other many-body approximations or from experiment.

\subsection{Multipole-Pad\'e representation}
\label{sec:MPA_representation}
Response functions obtained from first-principles calculations contain contributions from a large number of single-particle excitations, which are coupled through the Dyson equation in Eq.~\eqref{eq:dyson} and collectively give rise to structures such as plasmons. Despite this underlying complexity, the resulting frequency dependence of $Y(\omega)$ can often be described by a compact set of collective excitations with finite broadening. Multipole-Pad\'e approximants provide such an MPA representation, expressing the response in terms of a finite number of complex poles and associated spectral weights~\cite{Leon2021PRB,Leon2023PRB,Leon2025PRB,Leon2026PRB,Leon2026Arxiv}. In this work, we write the MPA representation in its time-ordered formulation as
\begin{equation}
Y_{\mathrm{MPA}}(\omega)=
\sum_{p=1}^{n_p}
\frac{r_p\Omega_p^2}
{\omega^2-\Omega_p^2},
\label{eq:mpa}
\end{equation}
where $n_p$ is the number of multipole contributions, $\Omega_p$ is a complex pole parameter, and $r_p$ is the corresponding weight.

We adopt the convention $\Re\,\Omega_p>0$ to label the pole with positive real part. For the time-ordered response, the imaginary part satisfies $\Im\,\Omega_p<0$. We additionally require
\begin{equation}
|\Im\,\Omega_p|<|\Re\,\Omega_p|,
\label{eq}
\end{equation}
so that the poles remain closer to the real-frequency axis than to the imaginary-frequency axis (see also Refs.~\cite{Leon2021PRB,Leon2023PRB,Leon2025PRB,Leon2026PRB}). 
The real and imaginary parts of $\Omega_p$ determine the excitation energy and broadening, respectively, while the weights are constrained to be real and positive, $r_p>0$.

\section{Machine-Learning Framework}
\label{sec:ML_framework}

\subsection{Inverse problem}
\label{sec:inverse_problem}
The MPA representation defines a forward mapping from a finite set of pole parameters to the frequency-dependent response. In the present work, we consider two forms of the inverse problem, depending on the information available in the input spectrum. The first uses the full complex response $Y(\omega)$, while the second uses only its imaginary part $\Im \,Y(\omega)$. The target of both inversions is the same set of pole parameters
\begin{eqnarray}
Y(\omega) &\longrightarrow
\left \{\Omega_p,r_p\right \}_{p=1}^{n_p}, 
\label{eq:inversion_full} \\
\Im\, Y(\omega) &\longrightarrow
\left \{\Omega_p,r_p\right \}_{p=1}^{n_p}.
\label{eq:inversion_Im}
\end{eqnarray}

The full-complex formulation is the most complete form and yields slightly more accurate results in the present study. The imaginary-only formulation is particularly relevant for processing experimental spectra where the imaginary component is the only one measured. Because the MPA parameters define the full complex response, their determination from $\Im Y(\omega)$ provides access to the corresponding real part without explicitly performing a Kramers-Kronig transformation over an extended frequency interval~\cite{Leon2024PRB,Leon2026JPCM}. The imaginary-only formulation therefore provides a route to reconstructing the full pole representation from experimentally accessible spectral information.

For both input representations, the network is trained to approximate the inverse mapping directly rather than determining the pole parameters through conventional nonlinear fitting. The physical constraints on the pole positions and residues, discussed in Sec.~\ref{sec:MPA_representation}, are explicitly incorporated into the output parameterization. 
To further assess the robustness of the inversion, we consider training ensembles with and without additional weak MPA contributions, which we refer to as ``noise'' throughout the remainder of this work. This results in four combinations: full-complex versus imaginary-only input, each trained with either clean spectra or spectra containing these additional weak contributions. These combinations allow us to separately assess the effects of reduced spectral information and weak spectral perturbations on the learned inversion.

\subsection{Synthetic ensemble}
\label{sec:ensemble}
To train the inverse mapping without relying on material-specific training data, we construct a synthetic ensemble of dynamical responses directly from the MPA representation in Eq.~\eqref{eq:mpa} with up to $n_p=5$ poles. For each $n_p=1,\ldots,5$, we generate $20000$ spectra by randomly sampling the pole parameters within prescribed ranges. Each spectrum is evaluated on a homogeneous frequency grid of $1000$ points in an interval of $100$~eV. The sampling ranges are
\begin{equation}
\begin{aligned}
r_p &\in [0.2,1],\\
\Re\, \Omega_p &\in [0.15,80]~\mathrm{eV},\\
\Im\, \Omega_p &\in [-10,-0.15]~\mathrm{eV}.
\end{aligned}
\label{eq}
\end{equation}
The lower bound on $r_p$ excludes poles with vanishing spectral weight, whose contribution to the response would become negligible and consequently difficult to identify. The lower bound on $\Re\,\Omega_p$ excludes poles with vanishing excitation energy, while the lower bound on $|\Im\Omega_p|$ avoids arbitrarily sharp poles and the associated poorly conditioned inversion. 

For spectra containing multiple poles, we additionally impose a Rayleigh-like criterion to distinguish neighboring excitations,
\begin{equation}
\Re\,\Omega_{p+1}-\Re\,\Omega_p >
\frac{|\Im\,\Omega_{p+1}+\Im\,\Omega_p|}{\alpha},
\label{eq}
\end{equation}
with $\alpha=1$. This condition prevents the training ensemble from containing pole pairs whose energy separation is too small relative to their combined broadening, following the
general notion of spectral resolvability~\cite{Jupp1998JMR,Robertson2013PASA}. The resulting ensemble therefore spans a controlled range of excitation energies, broadenings, spectral weights, and multipole complexity while retaining a well-defined correspondence between each response and its underlying pole parameters.

To assess the robustness of the inversion to weak spectral perturbations, we construct a second ensemble from the clean one by adding noise. The noise is generated by including $200$ additional MPA contributions with randomly sampled poles along the full 100~eV interval and small residues. The maximum residue of these noise poles is defined relative to the minimum physical residue as
\begin{equation}
\begin{aligned}
r_p^{\mathrm{noise}} &\in  [0, r_{\mathrm{max}}^{\mathrm{noise}}] \\
\eta_{\mathrm{noise}} &\equiv \frac{r_{\mathrm{max}}^{\mathrm{noise}}}{ r_{\min}}  = 10^{-3},
\label{eq:noise_strength}
\end{aligned}
\end{equation}
so that $r_{\mathrm{max}}^{\mathrm{noise}}=2\times10^{-4}$. These additional poles are not included in the target MPA representation and therefore act only as a controlled perturbation of the input spectra. We train separate models on the clean and noisy ensembles, allowing the effect of weak spectral noise on the learned inverse mapping to be assessed directly.

In practice, the resulting spectra are divided into training and validation sets using an 80\%  v.s. 20\% split for each value of $n_p$. The same partitioning is applied to both the clean and noisy ensembles, with the validation spectra excluded from optimization of the network parameters and used to monitor its performance during training.

\subsection{Neural-network architecture}
\label{sec:neural_network}
\begin{figure*}
    \centering
    \includegraphics[width=0.99\linewidth]{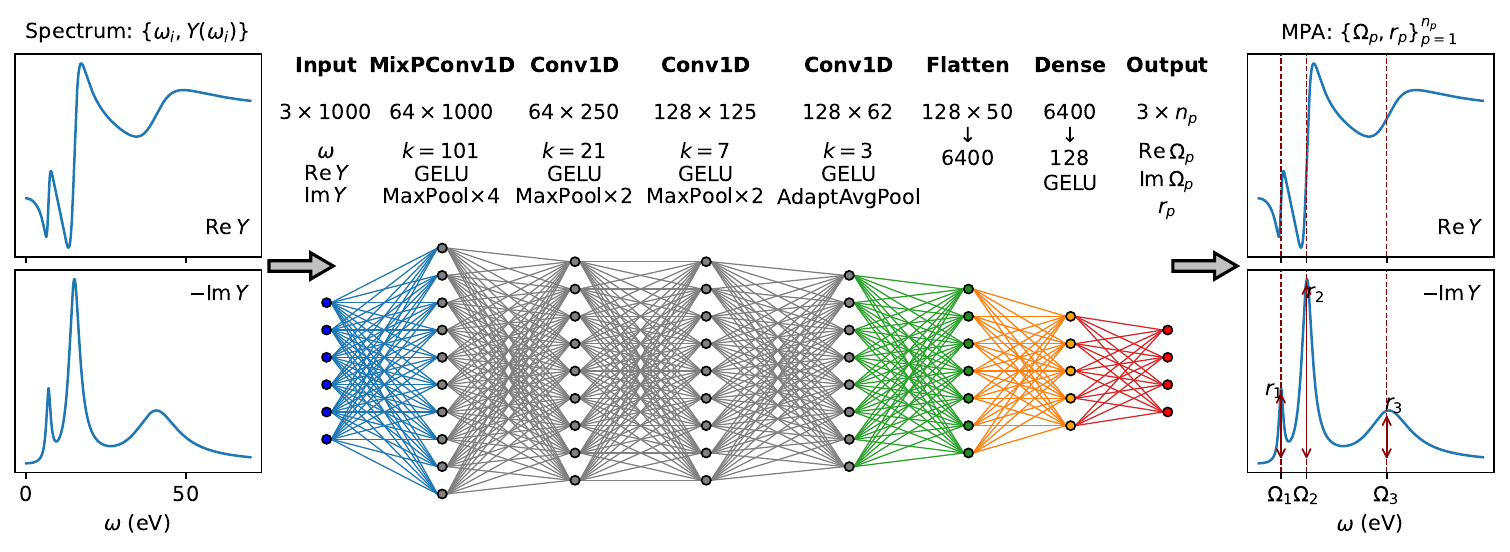}
    \caption{Schematic representation of the CNN architecture used to infer the MPA representation of the dielectric spectra. The input consists of the frequency grid and the real and imaginary components of the dielectric response, $\{\omega_i,\Re \,Y(\omega_i),\Im \,Y(\omega_i)\}_{i=1}^{n_{\omega}}$, sampled on $n_{\omega}=1000$ frequency points. The convolutional part of the network comprises four one-dimensional convolutional layers with kernel sizes $k=101$, $k=21$, $k=7$, and $k=3$, respectively, with GELU activations and max-pooling operations after the first three convolutional layers. An adaptive average-pooling operation reduces the resulting feature representation to $128\times50$, which is flattened and passed through a fully connected layer with 128 neurons and a GELU activation. The final linear layer outputs the $3n_p$ MPA parameters $\{\Re \,\Omega_p, \Im \,\Omega_p, r_p\}_{p=1}^{n_p}$. The spectra shown at the left and right illustrate the input response and its corresponding reconstruction from the predicted MPA parameters, respectively; the upper and lower panels show their real and imaginary components.}
    \label{fig:CNN}
\end{figure*}
The inverse mapping is implemented using a one-dimensional CNN with a fully connected regression head. The input is represented by separate channels for the frequency grid and the available components of $Y(\omega)$, with the specific input representation determined by the inverse problem defined in Eqs.~\eqref{eq:inversion_full} and~\eqref{eq:inversion_Im}. The convolutional layers progressively extract spectral features, with GELU activation functions applied after each convolutional layer and max-pooling operations used to reduce the frequency-axis dimensionality. The first convolutional layer uses a relatively large kernel to capture broad spectral structures, while subsequent three layers employ progressively smaller kernels to extract finer spectral features. An adaptive average-pooling operation then produces a fixed-size feature representation that is passed to the fully connected regression head.

For the architecture used in this work, the network contains around $1$ million trainable parameters, corresponding to approximately 4~MB of model weights in single-precision floating-point representation. The architecture was selected to achieve the desired accuracy while maintaining a compact model size and low inference cost. Therefore, it represents a practical choice that balances accuracy and computational cost rather than a fundamental limitation of the approach. The representational capacity can be increased by expanding the number of convolutional filters or fully connected neurons, while the framework can also accommodate different numbers of frequency samples and poles. In particular, increasing the number of poles requires only a corresponding increase in the dimensionality of the final output layer.

The final fully connected layer outputs $3n_p$ quantities corresponding to the real and imaginary parts of the poles and their associated residues, $(\Re,\Omega_p,\Im,\Omega_p,r_p)$. The physical constraints on these parameters are imposed through the output parameterization described in Sec.~\ref{sec:MPA_representation}. A schematic overview of the network architecture and the mapping from the input spectrum to the MPA parameters is shown in Fig.~\ref{fig:CNN}.

\subsection{Training objective}
\label{sec:training}
The network is trained using a loss function that combines the accuracy of the predicted MPA parameters with their ability to reproduce the input response. The parameter loss, the response-reconstruction loss and the total loss are defined as
\begin{eqnarray}
\mathcal{L}_{\mathrm{par}} &\equiv&
\frac{1}{n_p}
\sum_{p=1}^{n_p}
\left[ (\Delta\Omega_p)^2
+ w_r(\Delta r_p)^2
\right] 
\label{eq:loss_par} \\
\mathcal{L}_{\mathrm{res}} &\equiv&
\frac{1}{n_{\omega}} 
\sum_{i=1}^{n_{\omega}}
\left(\Delta Y(\omega_i)\right)^2 
\label{eq:loss_res} \\
\mathcal{L}_{\mathrm{tot}} & = & \mathcal{L}_{\mathrm{par}} + w_Y \mathcal{L}_{\mathrm{res}},
\label{eq:loss_tot}
\end{eqnarray}
where
\begin{equation}
    \begin{aligned}
     \Delta r_p &\equiv r_p^{\mathrm{pred}}-r_p^{\mathrm{true}} \\
    (\Delta\Omega_p)^2 &\equiv (\Re\, \Omega_p^{\mathrm{pred}}-\Re\, \Omega_p^{\mathrm{true}})^2 + \\
    &\,\,\,\,\,\,\,\, (\Im\, \Omega_p^{\mathrm{pred}}-\Im\, \Omega_p^{\mathrm{true}})^2 \\
   (\Delta Y)^2 & \equiv  (\Re\, Y_{\mathrm{pred}} - \Re\, Y_{\mathrm{true}})^2 + \\
   &\,\,\,\,\,\,\,\, (\Im\, Y_{\mathrm{pred}} - \Im\, Y_{\mathrm{true}})^2.
    \end{aligned}
\end{equation}
The weighting parameters $w_r$ and $w_Y$ control the relative contributions of the MPA residues and response-reconstruction terms to the total loss. They are fixed to $w_r=5$ and $w_Y=1$, which places greater emphasis on the poles than on the residues. 

This weighting is motivated by the structure of the inverse problem. The determination of the pole positions is nonlinear, whereas, once the pole positions are fixed, the corresponding residues can be obtained from a linear problem~\cite{Leon2021PRB,Leon2025PRB}. The selected weighting therefore prioritizes the more difficult nonlinear part of the inversion while retaining the response-reconstruction term to constrain the predicted parameters to reproduce the full spectral response.

Training is performed for a maximum of 300 epochs. Over this range, both the training and validation losses decrease by orders of magnitude and approach a saturation regime, with the two losses remaining close throughout the training. The network is therefore trained sufficiently far to reach the regime in which further optimization produces only marginal improvements, while avoiding the onset of overfitting.

\section{Results}
\label{sec:results}

\subsection{Validation on synthetic spectra}
\label{sec:validation}
\begin{figure*}
    \centering
        \includegraphics[width=0.99\textwidth]{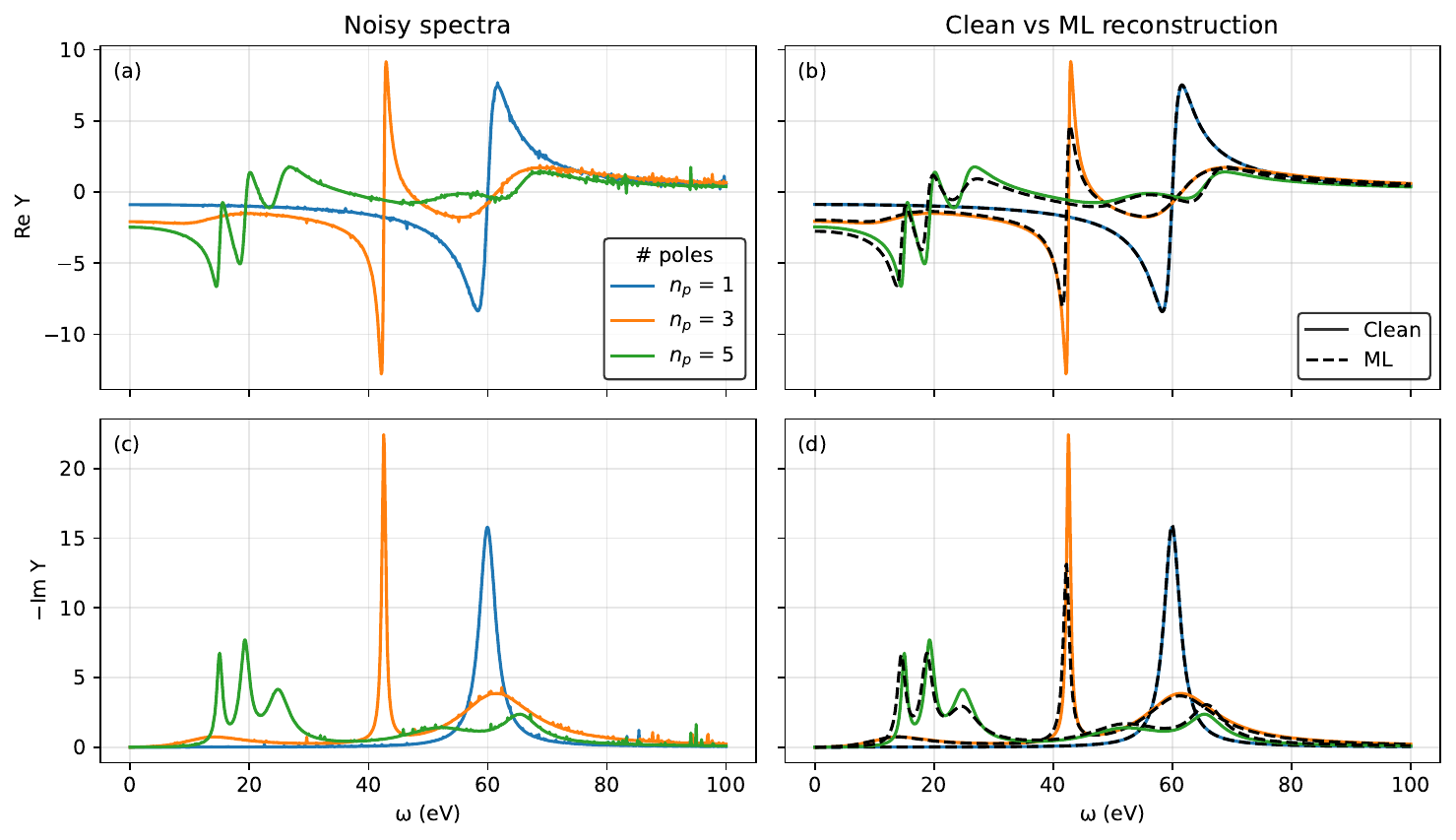}
    \caption{Examples of clean, noisy, and ML-reconstructed spectra generated with 3-5 poles. The real and imaginary parts are shown in the upper and lower rows, respectively. The left column (a, c)  shows the clean spectra and their noisy counterparts, while the right column (b, d) compares the clean spectra with the corresponding noise-trained ML reconstructions. Solid curves denote the clean spectra, and dashed black curves denote the ML reconstructions. Different colors correspond to different number of poles $n_p$.
    }
    \label{fig:synthetic_validation}
\end{figure*}
\begin{figure*}
    \centering
        \includegraphics[width=0.99\textwidth]{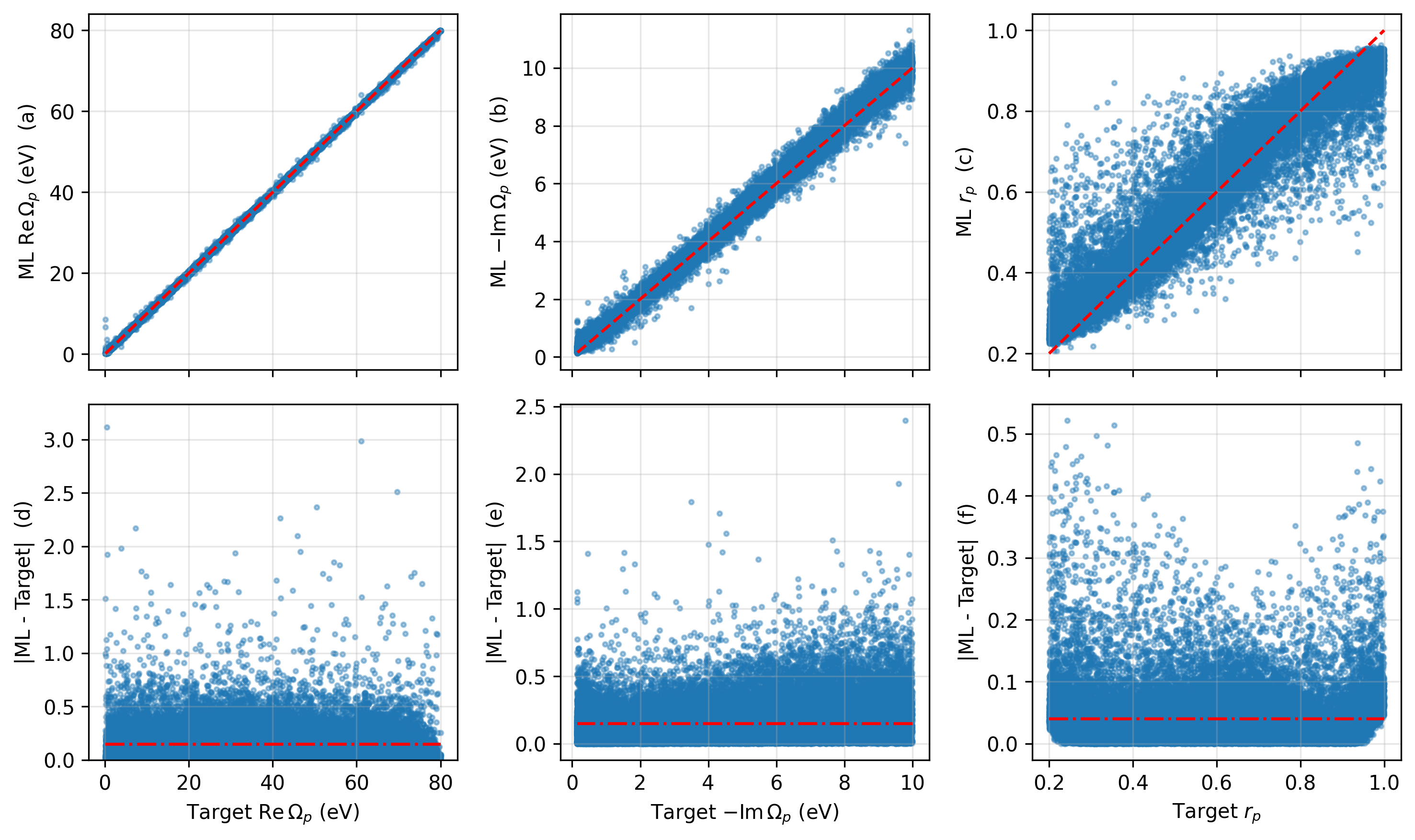}
    \caption{Comparison of the target and ML-predicted MPA parameters for all samples with $n_p=2$. The upper panels show the ML predictions as a function of the target values for $\Re\,\Omega_p$ (a), $\Im\,\Omega_p$ (b), and  $r_p$ (c). The dashed red lines indicate the identity relation. The corresponding absolute errors are shown in the lower panels (d-f), with horizontal dash-dotted lines indicating the mean errors. 
    }
    \label{fig:ensemble}
\end{figure*}
We first assess the ability of the CNN model to recover the MPA representation from the synthetic validation spectra. Fig.~\ref{fig:synthetic_validation} shows representative spectra generated from the MPA model for $n_p=1$, $3$, and $5$ poles. The real and imaginary components are shown separately in the upper and lower panels, respectively. In the left column, the clean spectra are compared with their noisy counterparts, illustrating the controlled spectral perturbations introduced by the additional weak contributions. In the right column, the clean spectra are compared with the responses reconstructed from the pole parameters predicted by the CNN model. The close correspondence between the reconstructed and target spectra demonstrates that the network can infer a compact MPA representation that accurately reproduces the frequency-dependent response.

For a more detailed assessment of the parameter predictions, Fig.~\ref{fig:ensemble} shows the ML results for the dataset with $n_p=2$, as the simplest case with spectral superposition. The close agreement between the predicted and target MPA parameters, together with the small absolute errors, demonstrates that the network can accurately recover the position, broadening, and residue of all the main poles. The same analysis is performed for all $n_p$ configurations considered in this work. 

\subsection{Model comparison}
\begin{figure*}
    \centering
        \includegraphics[width=0.99\textwidth]{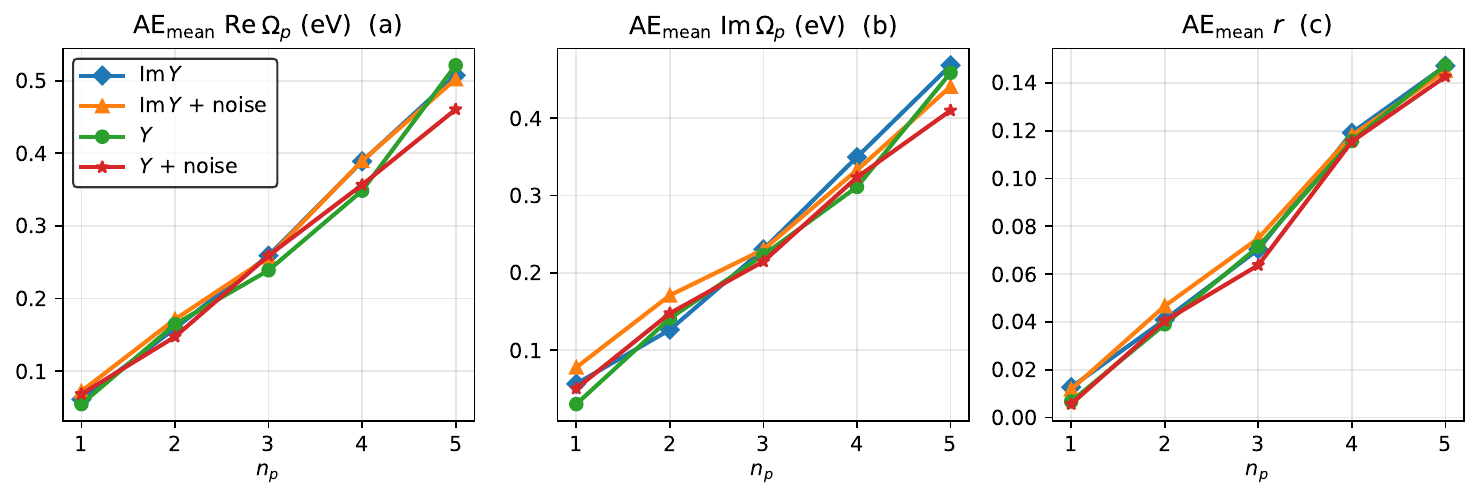}
    \caption{Comparison of the mean absolute error, $\text{AE}_{\text{mean}}$, of $\Re,\Omega_p$ (a), $\Im,\Omega_p$ (b), and $r_p$ (c) as a function of the number of poles $n_p$ for the four ML models, trained on $Y(\omega)$ or $\Im\,Y(\omega)$, with and without noise. 
    }
    \label{fig:error}
\end{figure*}
We next examine how the accuracy in the prediction of the MPA parameters changes with the complexity of the spectral response, for the four ML models. Fig.~\ref{fig:error} shows the mean absolute errors in $\Re,\Omega_p$, $\Im,\Omega_p$, and $r_p$ as a function of the number of poles, $n_p$, for models trained using the full complex response $Y(\omega)$ or only its imaginary part $\Im\,Y(\omega)$, with and without noise. In addition, 
Table~\ref{tab:pole_errors} summarizes the mean, median, and 95\% absolute errors 
of the MPA parameters 
for all values of $n_p$ and for all four ML models. 

\begin{table*}[htbp]
\centering
\begin{ruledtabular}
\begin{tabular}{llcccccccccccc}
& & \multicolumn{6}{c}{Training with noise} & \multicolumn{6}{c}{Training without noise} \\
\cline{3-8} \cline{9-14} \\[-6pt]
& & \multicolumn{3}{c}{Training on $Y$} & \multicolumn{3}{c}{Training on $\Im\,Y$} & \multicolumn{3}{c}{Training on $Y$} & \multicolumn{3}{c}{Training on $\Im\,Y$} \\
\cline{3-5} \cline{6-8} \cline{9-11} \cline{12-14} \\[-6pt]
Poles & Quantity & Mean & Median & 95\% & Mean & Median & 95\% & Mean & Median & 95\% & Mean & Median & 95\% \\
\hline\\[-8pt]
\multirow{4}{*}{$n_p=1$} & 
$\Re\,\Omega_p$ (eV) 
& 0.068 & 0.053 & 0.181  & 0.073 & 0.059 & 0.189    & 0.055 & 0.045 & 0.135  & 0.061 & 0.051 & 0.153 \\

& $\Im\,\Omega_p$ (eV) 
& 0.050 & 0.037 & 0.145  & 0.078 & 0.055 & 0.226    & 0.030 & 0.023 & 0.079  & 0.057 & 0.039 & 0.165 \\

& $r_p$           
& 0.006 & 0.004 & 0.015  & 0.012 & 0.009 & 0.033    & 0.007 & 0.004 & 0.023  & 0.013 & 0.008 & 0.040 \\


\hline\\[-8pt]
\multirow{4}{*}{$n_p=2$} & 
$\Re\,\Omega_p$ (eV) 
& 0.147 & 0.110 & 0.397  & 0.171 & 0.124 & 0.492    & 0.165 & 0.131 & 0.426  & 0.159 & 0.115 & 0.446 \\

& $\Im\,\Omega_p$ (eV) 
& 0.148 & 0.105 & 0.441  & 0.171 & 0.128 & 0.469    & 0.141 & 0.103 & 0.405  & 0.126 & 0.088 & 0.370 \\

& $r_p$           
& 0.040 & 0.029 & 0.118  & 0.047 & 0.031 & 0.154    & 0.039 & 0.027 & 0.116  & 0.041 & 0.027 & 0.134 \\


\hline\\[-8pt]
\multirow{4}{*}{$n_p=3$} & 
$\Re\,\Omega_p$ (eV) 
& 0.259 & 0.178 & 0.765  & 0.257 & 0.181 & 0.741    & 0.239 & 0.173 & 0.667  & 0.259 & 0.180 & 0.756 \\

& $\Im\,\Omega_p$ (eV) 
& 0.215 & 0.153 & 0.623  & 0.230 & 0.158 & 0.693    & 0.223 & 0.158 & 0.647  & 0.230 & 0.164 & 0.667 \\

& $r_p$           
& 0.064 & 0.045 & 0.194  & 0.075 & 0.053 & 0.225    & 0.072 & 0.050 & 0.217  & 0.070 & 0.048 & 0.218 \\


\hline\\[-8pt]
\multirow{4}{*}{$n_p=4$} & 
$\Re\,\Omega_p$ (eV) 
& 0.357 & 0.238 & 1.070  & 0.390 & 0.254 & 1.170    & 0.349 & 0.219 & 1.080  & 0.389 & 0.257 & 1.160 \\

& $\Im\,\Omega_p$ (eV) 
& 0.324 & 0.228 & 0.930  & 0.333 & 0.230 & 0.995    & 0.311 & 0.213 & 0.931  & 0.350 & 0.242 & 1.040 \\

& $r_p$           
& 0.115 & 0.090 & 0.308  & 0.118 & 0.092 & 0.314    & 0.116 & 0.089 & 0.310  & 0.119 & 0.095 & 0.312 \\


\hline\\[-8pt]
\multirow{4}{*}{$n_p=5$} & 
$\Re\,\Omega_p$ (eV) 
& 0.460 & 0.277 & 1.480  & 0.502 & 0.313 & 1.560    & 0.521 & 0.314 & 1.660  & 0.507 & 0.308 & 1.620 \\

& $\Im\,\Omega_p$ (eV) 
& 0.410 & 0.279 & 0.930  & 0.441 & 0.308 & 1.290    & 0.459 & 0.320 & 1.350  & 0.468 & 0.321 & 1.400 \\

& $r_p$           
& 0.143 & 0.118 & 0.360  & 0.145 & 0.121 & 0.363    & 0.147 & 0.124 & 0.363  & 0.147 & 0.121 & 0.372 \\


\end{tabular}
\end{ruledtabular}
\caption{Summary of error metrics for the MPA parameters 
obtained with the four ML models for all the used number of poles, $n_p$. 
We report the mean absolute error, the median absolute error and the 95\% absolute error.}
\label{tab:pole_errors}
\end{table*}

The four ML models show similar trends, with the mean, median, and 95\% absolute errors of the three MPA parameters increasing monotonically with $n_p$. This behavior reflects the increasing difficulty of resolving and assigning individual pole contributions as the number of overlapping excitations increases. The observed scaling suggests that the network architecture could be adapted to the spectral complexity, for example, by increasing its capacity as the number of poles increases. Nevertheless, such an adaptation is not required for the present application, as the present CNN architecture provides reasonable accuracy for all values of $n_p$ considered.

Moreover, the increase in the parameter errors does not necessarily correspond to a comparable increase in the error of the reconstructed response $Y(\omega)$. For spectra containing multiple poles, the individual contributions are superposed and weighted by their respective residues. A pole with a small spectral weight can therefore have a relatively large error in its position or broadening, while contributing only weakly to the total response. Conversely, the dominant poles determine much of the spectral structure and can be recovered more accurately even when weaker contributions are less precisely resolved. The parameter errors should therefore be considered together with the response-reconstruction errors when assessing the overall quality of the inversion.

The four ML models exhibit broadly similar error trends, with differences between training with and without noise that are generally small compared with the increase in error associated with increasing $n_p$. The model trained on the full complex $Y$ with noise shows a somewhat weaker increase in the mean error, particularly for the pole positions. However, this difference remains relatively small compared with the overall variation across the range of spectral complexities considered here. 

A slightly more systematic difference is observed between models trained using the full complex response $Y(\omega)$ and those trained using only $\Im\,Y(\omega)$. The former generally achieves lower parameter errors, indicating that the real part provides complementary information for determining the underlying pole structure. Nevertheless, the relatively close performance of the $\Im\,Y(\omega)$-only models is encouraging, as it shows that much of the information required to infer the MPA parameters is contained in the imaginary component alone. This result supports the use of the imaginary-only formulation for experimental applications in which $\Im\,Y(\omega)$ is the directly accessible spectral quantity.

\subsection{Application to first-principle spectra}
\begin{figure*}
    \centering
        \includegraphics[width=0.99\textwidth]{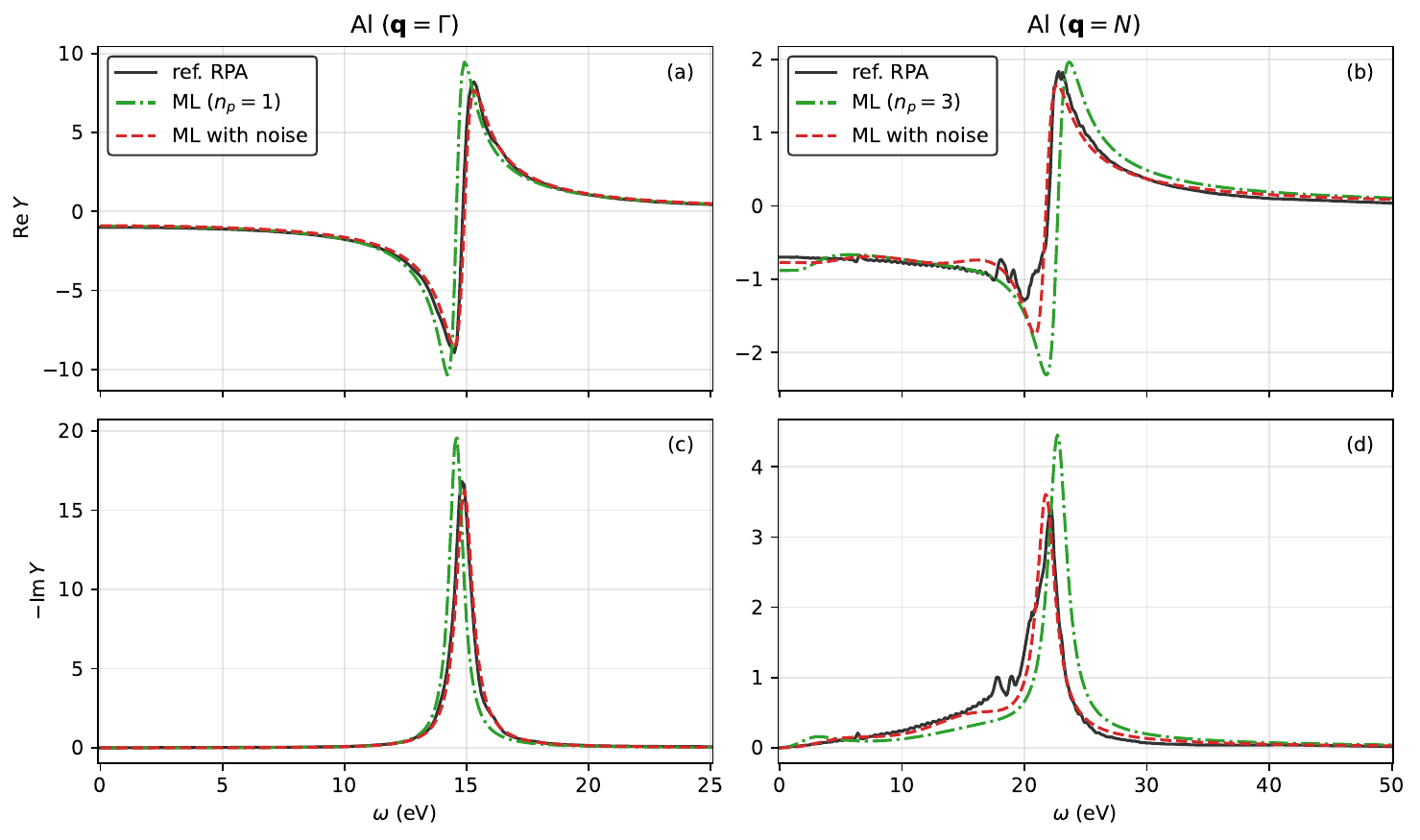}
    \caption{Application of the CNN model to RPA dielectric spectra of bulk Al for two momentum transfers $\q$. The real (a, b) and imaginary (c, d) parts of $Y(\omega)$ are shown in the upper and lower rows, respectively, for $\q=\Gamma$ (a, c) corresponding to the optical limit $\q=0$ and $\q=N$ (b, d), where $N$ denotes the Brillouin-zone boundary point considered here. The RPA spectra are shown together with the responses reconstructed from the MPA parameters predicted by CNN models trained with and without noise, with a number of poles $n_p=1$ for $\q=\Gamma$ and $n_p=3$ for $\q=N$.
    }
    \label{fig:Al}
\end{figure*}
\begin{figure*}
    \centering
    \includegraphics[width=0.99\textwidth]{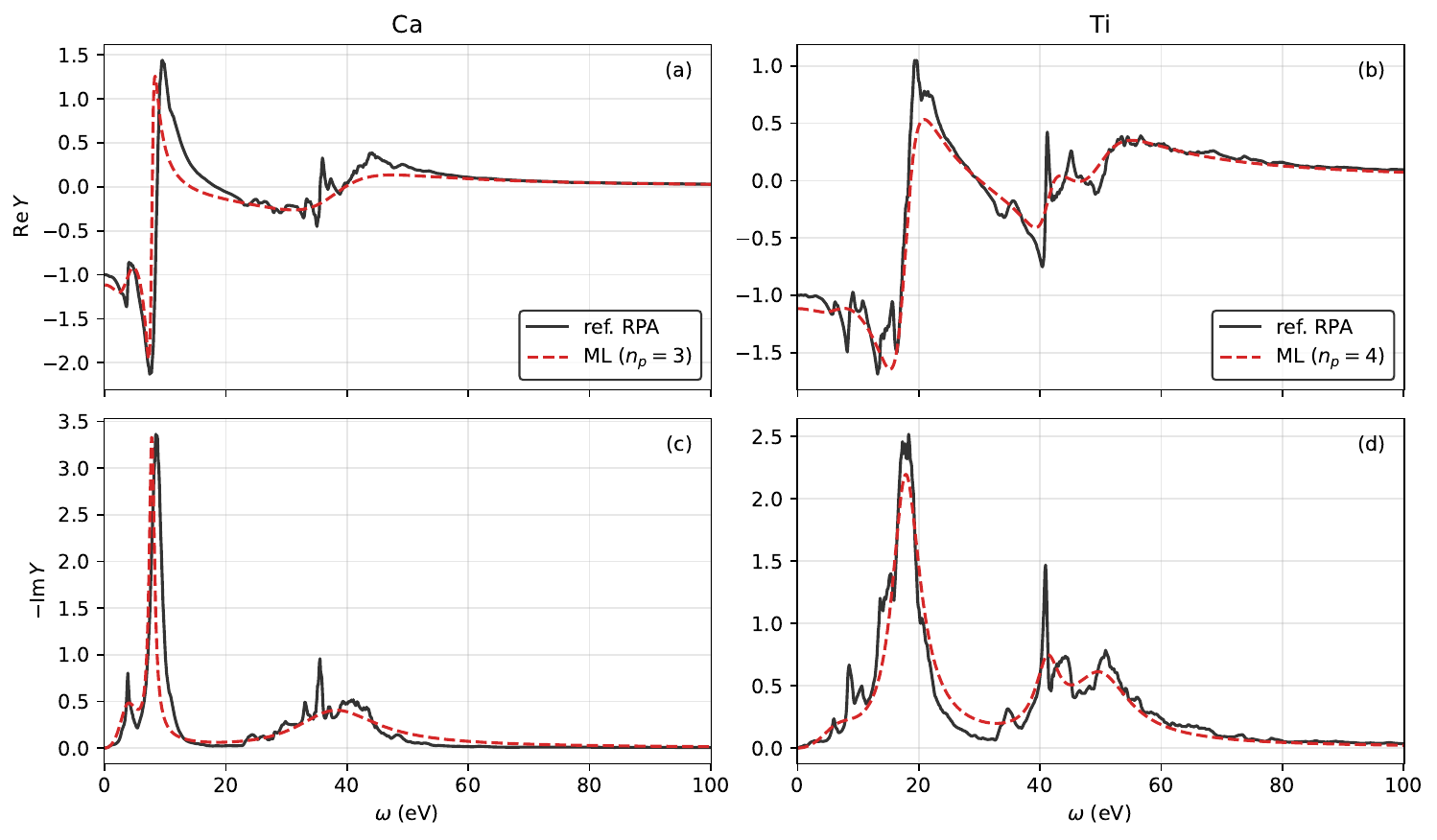}
    \caption{Application of the CNN model to RPA dielectric spectra of bulk Ca and Ti~\cite{Leon2026PRB} in the optical limit, $\q=0$. The real (a, b) and imaginary (c, d) parts of $Y(\omega)$ are shown in the upper and lower rows, respectively, for Ca (a, c) and Ti (b, d). In each case, the RPA spectrum is compared with the response reconstructed from the MPA parameters predicted by the best-performing CNN model, using $n_p=3$ for Ca and $n_p=4$ for Ti.}
    \label{fig:Ca_Ti}
\end{figure*}
We next assess the ability of the CNN to generalize from the synthetic training ensemble to dielectric responses obtained from first-principles calculations. Figure~\ref{fig:Al} shows RPA spectra of bulk Al for two momentum transfers, $\q=\Gamma$ and $\q=N$, together with the corresponding responses reconstructed from the MPA parameters predicted by CML models trained with and without noise. 
For $\q=\Gamma$, the main spectral structure is captured using $n_p=1$, while $n_p=3$ poles are used for $\q=N$. In both cases, the CNN reconstruction closely follows the RPA response over the frequency range considered, including the main features in both the real and imaginary components. The model trained with noise provides the closer reconstruction in these cases, consistent with the improved behavior observed for this model in the synthetic validation. More importantly, the close agreement with the first-principles spectra demonstrates that the CNN can transfer the inverse mapping learned from synthetic MPA responses to dielectric responses that are not part of the training ensemble.

A more demanding test is provided by bulk Ca and Ti, shown in Fig.~\ref{fig:Ca_Ti}. Their dielectric responses exhibit more complex spectral structures than those considered for Al, requiring $n_p=3$ and $4$ poles, respectively, to accurately capture the main spectral structures. Despite this increased complexity, the ML predictions reproduce the first-principles responses with high accuracy for both materials. The spectra also contain secondary structures that would require additional poles for a more detailed representation, and may not satisfy some of the conditions imposed in the construction of the synthetic ensemble in Sec.~\ref{sec:ensemble}. Nevertheless, the trained CNN provides a good reconstruction of the dominant spectral response within the present pole representation. The successful reconstruction of these distinct first-principles spectra demonstrates the general applicability of the proposed approach and shows that the synthetic training ensemble captures features of the inverse mapping that transfer beyond the spectra used to construct the training data.

\subsection{Application to experimental spectra}
\begin{figure*}
    \centering
    \includegraphics[width=0.99\textwidth]{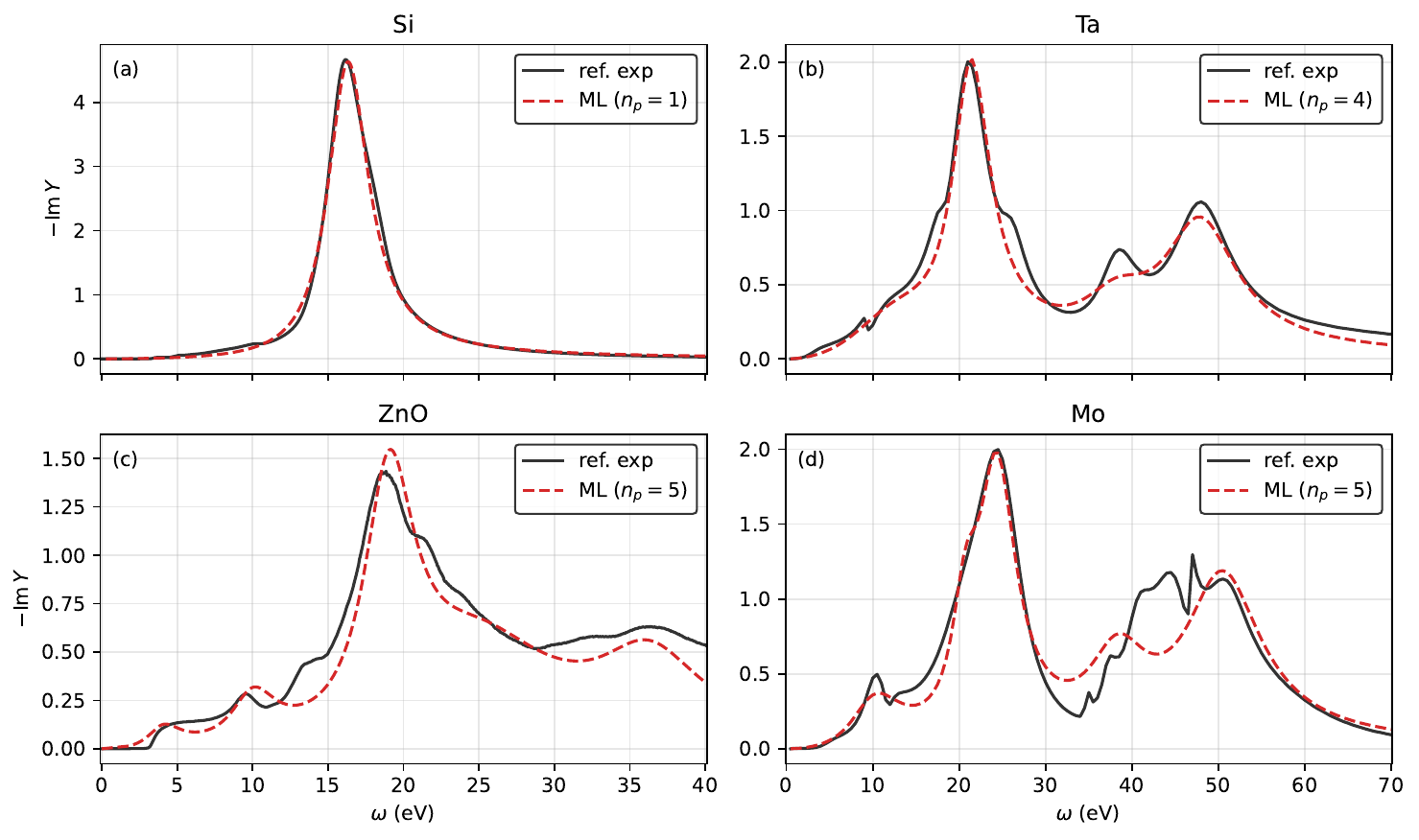}
    \caption{Application of the CNN model to experimental dielectric spectra of bulk Si~\cite{Franta2017ASS}, ZnO~\cite{Leon2024PRB}, Ta, and Mo~\cite{Werner2009JPCRD}. The plots show $\Im\,Y(\omega)$ for Si (a), Ta (b), ZnO (c), and Mo (d). The experimental spectra are compared with the corresponding responses reconstructed from the MPA parameters predicted by the best-performing CNN model, using $n_p=1$ for Si, $n_p=4$ for Ta and $n_p=5$ for ZnO and Mo.}
    \label{fig:Si_ZnO_Ta_Mo}
\end{figure*}
We finally assess the performance of the CNN model on experimental dielectric spectra spanning different material classes and spectral characteristics. Figure~\ref{fig:Si_ZnO_Ta_Mo} shows the imaginary part of the inverse dielectric response, $\Im\,Y(\omega)$, for bulk Si, ZnO, Ta, and Mo. The Si spectrum is taken from Ref.~\cite{Franta2017ASS}, the ZnO spectrum from Ref.~\cite{Leon2024PRB}, and the Ta and Mo spectra from Ref.~\cite{Werner2009JPCRD}. These systems provide stringent tests of the learned inversion because they exhibit spectral features that extend beyond those represented in the synthetic training ensemble.

The spectrum of Si, an indirect-band-gap semiconductor, exhibits a relatively simple response that is accurately captured using a single pole, $n_p=1$. The metallic systems Ta and Mo display substantially richer spectral structures. For Ta, the main spectral features are accurately reproduced with $n_p=4$ poles, while the response of Mo is reasonably described using $n_p=5$. Additional structures are visible in the Mo spectrum that would require a larger number of poles for a more detailed representation. Nevertheless, the agreement for the dominant features demonstrates that the CNN can extract a useful MPA representation from experimental spectra with considerably richer structures than those used during training.

The experimental spectrum of ZnO provides a more challenging test. As a direct-band-gap semiconductor, ZnO exhibits a pronounced spectral onset in addition to several distinct peaks. Such an onset is not represented in the synthetic ensemble, which consists of spectra generated from a finite set of discrete poles, and therefore cannot be expected to be reproduced with the same accuracy as the plasmonic pole structures. The experimental spectrum also contains a finite level of measurement noise, which differs from the controlled weak MPA contributions introduced as noise in the synthetic ensemble. In addition, obtaining an accurate low-energy tail from experimental data can be sensitive to the treatment of the background and other post-processing procedures~\cite{Leon2024PRB}. In the present case, the measured tail is incomplete, which may further limit the accuracy of the reconstructed response. The ZnO spectrum therefore provides an example in which several aspects of the experimental response are not represented in the training ensemble.

Shoulder structures provide another example of spectral features that are not explicitly included in the synthetic ensemble. This is particularly apparent for Ta, where the main peaks are well reproduced by the CNN reconstruction, while two weaker shoulder structures around the main peak remain unresolved. Similar secondary structures are also visible in the Mo and ZnO spectra. These deviations illustrate that the accuracy of the inversion is ultimately related to how well the spectral characteristics encountered in the target data are represented by the synthetic training ensemble.

The experimental spectra considered here were selected to provide stringent tests of the method, spanning semiconductors and metals and including spectral structures that extend beyond the characteristics of the synthetic ensemble. Despite these differences, the CNN model provides a good reconstruction of the dominant spectral response for all four materials. The results therefore provide a further demonstration that a model trained entirely on synthetic spectra can be applied to experimental dielectric responses without material-specific training.

\section{Conclusions}
\label{sec:conclusions}
We have demonstrated that the inverse mapping between a dielectric response and its underlying plasmonic structure can be learned directly from synthetic spectra generated from an analytical multipole-Pad\'e representation. The designed CNN does not require training on spectra associated with specific materials, but instead learns the relation between the spectral response and the parameters of its underlying MPA representation. The resulting model can therefore be transferred to first-principles and experimental dielectric spectra that are not represented in the training data.

The results also show that the complexity of the inverse problem increases with the number of overlapping excitations. As the number of poles increases, the uncertainty in the individual pole parameters becomes larger, while the reconstructed response can remain accurate because weak contributions have a limited influence on the total spectrum. The comparison between models trained with the full complex response and with only its imaginary part further shows that the imaginary component contains enough information to recover the underlying MPA structure, despite the generally improved accuracy obtained from the full response.

The application to experimental spectra provides an important test of the scope of the present approach. The model successfully reproduces the dominant structures of spectra from both semiconductors and metals, including responses that are substantially more complex than the synthetic training examples. On the other hand, the experimental results expose limitations associated with features that are not represented in the current synthetic ensemble, including spectral onsets, shoulder structures, incomplete spectral tails, and experimental noise that differs from the controlled perturbations used during training. These examples demonstrate that the present implementation should be viewed as a proof of feasibility rather than as a representation optimized for all possible experimental spectral features.

An important advantage of the approach is that these limitations can be addressed systematically. The synthetic ensemble can be extended to include additional spectral structures, closer and more strongly overlapping poles, larger numbers of poles, and experimentally relevant forms of noise and background variation. Increasing the diversity of the ensemble would naturally require corresponding increases in the training-set size and, where necessary, in the capacity of the neural network. The present results therefore establish a framework in which the accuracy and range of applicability of the inversion can be improved by systematically expanding the underlying synthetic representation.

Overall, the study demonstrates the feasibility of using physics-constrained synthetic training to automate the extraction of plasmonic pole representations from complex dielectric spectra. The ability to apply the same learned inverse mapping across different materials and to spectra obtained from both first-principles calculations and experiment provides a basis for high-throughput analysis of collective electronic excitations, while the controlled construction of the synthetic ensemble offers a direct route for extending the method to increasingly complex spectral responses.

\section*{Acknowledgments}
This work was partially supported by the Research Council of Norway through the DOPED project (335022). 

\section*{Data Availability}
The data generated in this article are openly available through the corresponding author.

%
%
\bibliography{biblio}

@book{martin2016book,
  author={Martin,Richard M. and Reining,Lucia and Ceperley,David M.},
  title={Interacting Electrons},
  year=2016,
  publisher={Cambridge University Press},
  address={Cambridge}
}

@book{Egerton2011book,
  title={Electron Energy-Loss Spectroscopy in the Electron Microscope},
  author={Egerton, RF},
  year={2011},
  publisher={Springer},
  doi={10.1007/978-1-4419-9583-4}
}

@book{Schulke2007book,
  author={Schülke, Winfried},
  title={Electron dynamics by inelastic X-ray scattering},
  year=2007,
  volume={7},
  publisher={OUP Oxford},
}

@article{Werner2009JPCRD,
author = {Werner, Wolfgang~S.~M.  and Glantschnig, Kathrin  and {Ambrosch-Draxl},Claudia},
title = {Optical Constants and Inelastic Electron-Scattering Data for 17 Elemental Metals},
journal = {Journal of Physical and Chemical Reference Data},
volume = {38},
number = {4},
pages = {1013-1092},
year = {2009},
doi = {10.1063/1.3243762},
}

@article{Franta2017ASS,
title = {Temperature-dependent dispersion model of float zone crystalline silicon},
journal = {Applied Surface Science},
volume = {421},
pages = {405-419},
year = {2017},
note = {7th International Conference on Spectroscopic Ellipsometry},
issn = {0169-4332},
doi = {https://doi.org/10.1016/j.apsusc.2017.02.021},
author = {Daniel Franta and Adam Dubroka and Chennan Wang and Angelo Giglia and Jir{í} Voh{á}nka and Pavel Franta and Ivan Ohl{í}dal},
}

@article{Leon2026JPCM,
	author={Leon, Dario A. and Berland, Kristian},
	title={Momentum- and frequency-resolved collective electronic excitations in solids: insights from spectroscopy and first-principles calculations},
	journal={Journal of Physics: Condensed Matter},
	doi = {https://doi.org/10.1088/1361-648X/ae62e7},
	year={2026},
    volume = {38},
    issue = {18},
    pages = {183001},
}

@misc{Leon2026Arxiv,
   author = {Leon, Dario A. and Cardoso, Claudia and Berland, Kristian},
   title = {Plasmonic properties and correlation energies from a compact multipole representation of the dielectric response in 2D metals},
   howpublished = {\url{https://arxiv.org/abs/2606.12274}},
   year = {2026}
}

@article{Leon2026PRB,
  title = {Bulk plasmons in elemental metals},
  author = {Leon, Dario A. and Cardoso, Claudia and Berland, Kristian},
  journal = {Phys. Rev. B},
  volume = {113},
  issue = {12},
  pages = {125138},
  numpages = {13},
  year = {2026},
  month = {Mar},
  publisher = {American Physical Society},
  doi = {https://doi.org/10.1103/w51n-p91p},
}

@article{Leon2021PRB,
  title = {Frequency dependence in $GW$ made simple using a multipole approximation},
  author = {Leon, Dario A. and Cardoso, Claudia and Chiarotti, Tommaso and Varsano, Daniele and Molinari, Elisa and Ferretti, Andrea},
  journal = {Phys. Rev. B},
  volume = {104},
  issue = {11},
  pages = {115157},
  numpages = {16},
  year = {2021},
  doi = {https://doi.org/10.1103/PhysRevB.104.115157},
}

@article{Leon2023PRB,
  title = {Efficient full frequency GW for metals using a multipole approach for the dielectric screening},
  author = {Leon, Dario A. and Ferretti, Andrea and Varsano, Daniele and Molinari, Elisa and Cardoso, Claudia},
  journal = {Phys. Rev. B},
  volume = {107},
  issue = {15},
  pages = {155130},
  numpages = {15},
  year = {2023},
  month = {Apr},
  publisher = {American Physical Society},
  doi = {10.1103/PhysRevB.107.155130},
  url = {https://link.aps.org/doi/10.1103/PhysRevB.107.155130}
}

@article{Leon2024PRB,
  title = {Unraveling many-body effects in ZnO: Combined study using momentum-resolved electron energy-loss spectroscopy and first-principles calculations},
  author = {Leon, Dario A. and Elgvin, Cana and Nguyen, Phuong Dan and Prytz, \O{}ystein and Hage, Fredrik S. and Berland, Kristian},
  journal = {Phys. Rev. B},
  volume = {109},
  issue = {11},
  pages = {115153},
  numpages = {17},
  year = {2024},
  month = {Mar},
  publisher = {American Physical Society},
  doi = {10.1103/PhysRevB.109.115153},
  url = {https://link.aps.org/doi/10.1103/PhysRevB.109.115153}
}

@article{Leon2025PRB,
  title = {Spectral properties from an efficient analytical representation of the $GW$ self-energy within a multipole approximation},
  author = {Leon, Dario A. and Berland, Kristian and Cardoso, Claudia},
  journal = {Phys. Rev. B},
  volume = {111},
  issue = {19},
  pages = {195147},
  numpages = {16},
  year = {2025},
  month = {May},
  publisher = {American Physical Society},
  doi = {10.1103/PhysRevB.111.195147},
  url = {https://link.aps.org/doi/10.1103/PhysRevB.111.195147}
}

@article{Guandalini2024PRB,
  title = {Efficient {GW} calculations via interpolation of the screened interaction in momentum and frequency space: The case of graphene},
  author = {Guandalini, Alberto and Leon, Dario A. and D'Amico, Pino and Cardoso, Claudia and Ferretti, Andrea and Rontani, Massimo and Varsano, Daniele},
  journal = {Phys. Rev. B},
  volume = {109},
  issue = {7},
  pages = {075120},
  numpages = {15},
  year = {2024},
  month = {Feb},
  publisher = {American Physical Society},
  doi = {10.1103/PhysRevB.109.075120},
}

@article{Bonacci2026JCTC,
    author = {Bonacci, M. and Leon, D. A. and Spallanzani, N. and Molinari, E. and Varsano, D. and Ferretti, A. and Cardoso, C.},
    title = {Benchmarking the Plasmon-Pole and Multipole Approximations
in the Yambo Code Using the GW100 Data Set},
    journal = {Journal of Chemical Theory and Computation},
    volume = {22},
    number = {13},
    pages = {6718-6730},
    year = {2026},
    month = {07},
    issn = {1549-9618},
    doi = {10.1021/acs.jctc.6c00239},
    url = {https://doi.org/10.1021/acs.jctc.6c00239},
}

@article{Soininen2003JPCM,
   author = {J.~A. Soininen and J.~J. Rehr and Eric~L. Shirley},
   title = {Electron self-energy calculation using a general multi-pole approximation},
   journal = {J. Phys.: Condens. Matter},
   volume = {15},
   number = {},
   pages = {2573-2586},
   year = {2003},
   doi = {https://doi.org/10.1088/0953-8984/15/17/312}
}

@article{Lee1996PRB,
   author = {Keun-Ho Lee and K.~J. Chang},
   title = {Analytic continuation of the dynamic response function using an {N}-point {P}ad\'e approximant},
   journal = {Phys. Rev. B},
   volume = {54},
   number = {12},
   pages = {R8285-R8288},
   year = {1996},
   doi = {https://doi.org/10.1103/physrevb.54.r8285}
}

@article{Jin1999PRB,
   author = {Young-Gu Jin and K.~J. Chang},
   title = {Dynamic response function and energy-loss spectrum for {L}i using an {N}-point {P}ad\'e approximant},
   journal = {Phys. Rev. B},
   volume = {59},
   number = {23},
   pages = {R8285-R8288},
   year = {1999},
   doi = {https://doi.org/10.1103/PhysRevB.59.14841}
}

@Article{Golze2019FC,
author={Golze, Dorothea  and Dvorak, Marc  and Rinke, Patrick },     
title={The GW Compendium: A Practical Guide to Theoretical Photoemission Spectroscopy},   
journal={Frontiers in Chemistry},   
volume={7},
year={2019},
doi={https://doi.org/10.3389/fchem.2019.00377},
issn={2296-2646},
}

@Article{Ward2016npjCompMat,
author={Ward, Logan
and Agrawal, Ankit
and Choudhary, Alok
and Wolverton, Christopher},
title={A general-purpose machine learning framework for predicting properties of inorganic materials},
journal={npj Computational Materials},
year={2016},
month={Aug},
day={26},
volume={2},
number={1},
pages={16028},
issn={2057-3960},
doi={10.1038/npjcompumats.2016.28},
url={https://doi.org/10.1038/npjcompumats.2016.28}
}

@Article{Chen2020npjCompMat,
author={Chen, Lihua
and Kim, Chiho
and Batra, Rohit
and Lightstone, Jordan P.
and Wu, Chao
and Li, Zongze
and Deshmukh, Ajinkya A.
and Wang, Yifei
and Tran, Huan D.
and Vashishta, Priya
and Sotzing, Gregory A.
and Cao, Yang
and Ramprasad, Rampi},
title={Frequency-dependent dielectric constant prediction of polymers using machine learning},
journal={npj Computational Materials},
year={2020},
month={May},
day={21},
volume={6},
number={1},
pages={61},
issn={2057-3960},
doi={10.1038/s41524-020-0333-6},
url={https://doi.org/10.1038/s41524-020-0333-6}
}

@article{Grunert2024PRM,
  title = {Deep learning of spectra: Predicting the dielectric function of semiconductors},
  author = {Grunert, Malte and Gro\ss{}mann, Max and Runge, Erich},
  journal = {Phys. Rev. Mater.},
  volume = {8},
  issue = {12},
  pages = {L122201},
  numpages = {6},
  year = {2024},
  month = {Dec},
  publisher = {American Physical Society},
  doi = {10.1103/PhysRevMaterials.8.L122201},
  url = {https://link.aps.org/doi/10.1103/PhysRevMaterials.8.L122201}
}

@article{Dong2021CS,
    author = {Dong, Sijia S. and Govoni, Marco and Galli, Giulia},
    title = {Machine learning dielectric screening for the simulation of excited state properties of molecules and materials},
    journal = {Chemical Science},
    volume = {12},
    number = {13},
    pages = {4970-4980},
    year = {2021},
    month = {04},
    issn = {2041-6520},
    doi = {10.1039/d1sc00503k},
    url = {https://doi.org/10.1039/d1sc00503k},
}

@article{Borlido2026PRM,
  title = {Screening for wide-gap, high-refractive-index materials via frequency-dependent dielectric function prediction aided by convolutional autoencoders},
  author = {Borlido, Pedro and Nogueira, Fernando},
  journal = {Phys. Rev. Mater.},
  volume = {10},
  issue = {6},
  pages = {064606},
  numpages = {12},
  year = {2026},
  month = {Jun},
  publisher = {American Physical Society},
  doi = {10.1103/dg72-zpqn},
  url = {https://link.aps.org/doi/10.1103/dg72-zpqn}
}

@article{Zhang2020PRB,
  title = {Deep neural network for the dielectric response of insulators},
  author = {Zhang, Linfeng and Chen, Mohan and Wu, Xifan and Wang, Han and E, Weinan and Car, Roberto},
  journal = {Phys. Rev. B},
  volume = {102},
  issue = {4},
  pages = {041121(R)},
  numpages = {6},
  year = {2020},
  month = {Jul},
  publisher = {American Physical Society},
  doi = {10.1103/PhysRevB.102.041121},
  url = {https://link.aps.org/doi/10.1103/PhysRevB.102.041121}
}

@Article{Gleason2024npjCompMat,
author={Gleason, Samuel P.
and Lu, Deyu
and Ciston, Jim},
title={Prediction of the Cu oxidation state from EELS and XAS spectra using supervised machine learning},
journal={npj Computational Materials},
year={2024},
month={Sep},
day={17},
volume={10},
number={1},
pages={221},
issn={2057-3960},
doi={10.1038/s41524-024-01408-1},
url={https://doi.org/10.1038/s41524-024-01408-1}
}

@Article{Shimano2023SciRep,
author={Shimano, Yuho
and Kutana, Alex
and Asahi, Ryoji},
title={Machine learning and atomistic origin of high dielectric permittivity in oxides},
journal={Scientific Reports},
year={2023},
month={Dec},
day={14},
volume={13},
number={1},
pages={22236},
issn={2045-2322},
doi={10.1038/s41598-023-49603-2},
url={https://doi.org/10.1038/s41598-023-49603-2}
}

@article{Grumet2024JPCC,
    author = {Grumet, Manuel and von Scarpatetti, Clara and Bucko, Tomás and Egger, David A.},
    title = {Delta Machine Learning for Predicting Dielectric Properties
and Raman Spectra},
    journal = {The Journal of Physical Chemistry C},
    volume = {128},
    number = {15},
    pages = {6464-6470},
    year = {2024},
    month = {04},
    issn = {1932-7447},
    doi = {10.1021/acs.jpcc.4c00886},
    url = {https://doi.org/10.1021/acs.jpcc.4c00886},
}

@article{Mizoguchi2026Micro,
    author = {Mizoguchi, Teruyasu},
    title = {Data-driven ELNES/XANES analysis: predicting spectra, unveiling structures and quantifying properties},
    journal = {Microscopy},
    volume = {75},
    number = {2},
    pages = {164-184},
    year = {2026},
    month = {04},
    issn = {2050-5701},
    doi = {10.1093/jmicro/dfaf038},
    url = {https://doi.org/10.1093/jmicro/dfaf038},
}

@article{Ibrahim2024AMI,
    author = {Ibrahim, Akram and Ataca, Can},
    title = {Prediction of Frequency-Dependent
Optical Spectrum
for Solid Materials: A Multioutput and Multifidelity Machine Learning
Approach},
    journal = {ACS Applied Materials \& Interfaces},
    volume = {16},
    number = {31},
    pages = {41145-41156},
    year = {2024},
    month = {07},
    issn = {1944-8244},
    doi = {10.1021/acsami.4c07328},
    url = {https://doi.org/10.1021/acsami.4c07328},
}

@article{Stein2019CS,
    author = {Stein, Helge S. and Guevarra, Dan and Newhouse, Paul F. and Soedarmadji, Edwin and Gregoire, John M.},
    title = {Machine learning of optical properties of materials – predicting spectra from images and images from spectra},
    journal = {Chemical Science},
    volume = {10},
    number = {1},
    pages = {47-55},
    year = {2019},
    month = {01},
    issn = {2041-6520},
    doi = {10.1039/c8sc03077d},
    url = {https://doi.org/10.1039/c8sc03077d},
}

@article{Anker2023CS,
    author = {Anker, Andy S. and Butler, Keith T. and Selvan, Raghavendra and Jensen, Kirsten M. O.},
    title = {Machine learning for analysis of experimental scattering and spectroscopy data in materials chemistry},
    journal = {Chemical Science},
    volume = {14},
    number = {48},
    pages = {14003-14019},
    year = {2023},
    month = {12},
    issn = {2041-6520},
    doi = {10.1039/d3sc05081e},
    url = {https://doi.org/10.1039/d3sc05081e},
}

@article{Penfold2024MLsciandtech,
doi = {10.1088/2632-2153/ad5074},
url = {https://doi.org/10.1088/2632-2153/ad5074},
year = {2024},
month = {jun},
publisher = {IOP Publishing},
volume = {5},
number = {2},
pages = {021001},
author = {Penfold, Thomas and Watson, Luke and Middleton, Clelia and David, Tudur and Verma, Sneha and Pope, Thomas and Kaczmarek, Julia and Rankine, Conor},
title = {Machine-learning strategies for the accurate and efficient analysis of x-ray spectroscopy},
journal = {Machine Learning: Science and Technology},
}

@article{del-Pozo-Bueno2023Um,
title = {Comparative of machine learning classification strategies for electron energy loss spectroscopy: Support vector machines and artificial neural networks},
journal = {Ultramicroscopy},
volume = {253},
pages = {113828},
year = {2023},
issn = {0304-3991},
doi = {https://doi.org/10.1016/j.ultramic.2023.113828},
url = {https://www.sciencedirect.com/science/article/pii/S0304399123001456},
author = {Daniel del-Pozo-Bueno and Demie Kepaptsoglou and Francesca Peiró and Sònia Estradé},
}

@article{Jupp1998JMR,
title = {Precision in Estimating the Frequency Separation between Spectral Lines},
journal = {Journal of Magnetic Resonance},
volume = {135},
number = {1},
pages = {23-29},
year = {1998},
issn = {1090-7807},
doi = {https://doi.org/10.1006/jmre.1998.1538},
url = {https://www.sciencedirect.com/science/article/pii/S1090780798915385},
author = {P.E Jupp and K.D.M Harris and A.E Aliev},
}

@article{Robertson2013PASA, 
title={Quantifying Resolving Power in Astronomical Spectra}, 
volume={30}, 
DOI={10.1017/pasa.2013.26}, 
journal={Publications of the Astronomical Society of Australia}, 
author={Robertson, J. Gordon}, 
year={2013}, 
pages={e048}
}

\end{document}